# Prospects for the implementation of an affordable VR/AR-content management tool for Learning Management Systems

Anastasia Grigoreva [1] and Stanislav Grigorev [1*]

[1] Irkutsk National Research Technical University; svg@istu.edu

* Correspondence: svg@istu.edu;

**ABSTRACT**

The article discusses the prospects for introducing into educational practice the designer of electronic training courses based on virtual and augmented reality technologies for LMS Moodle. The requirements for the functions, interface, appearance of the module-designer being developed, the formation of VR/AR-content in terms of its use by unprepared users, such as teachers and developers of training courses, are formulated.

**Keywords:** online learning; learning management system; Moodle; virtual reality; augmented reality; software development.

Virtual and augmented reality is increasingly mentioned in the media in the context of computer games, advertising or social services. However, the possibilities of VR (virtual reality) and AR (augmented reality) are much wider and at the moment these technologies are already beginning to be actively used in a wide variety of applications.

The fundamental difference between VR and AR is that VR blocks the real world and immerses the user in the digital universe, while AR adds elements of the digital world to the real one [1].

The fact that over the past 5-10 years the percentage of virtual reality use elements in education has increased due to the spread and availability on the market of appropriate technical means (VR glasses, adapters, set-top boxes, etc.), increased computing capabilities portable devices, expanding network and Internet traffic and increasing the speed of information transfer [2]. All this contributes to the rapid introduction of VR/AR technology not only in education, but in our entire life.

One of the new and promising areas of application of VR and AR is pre-university and higher education. For learning, both virtual and augmented reality can play a useful role and increase the effectiveness of the perception of educational content. E-learning, even in the form of webinars, has a number of significant drawbacks, such as the effect of distraction or "simulation" of presence.

Deeper immersion in any environment, including educational environment, can be guaranteed by the visual and vestibular illusion that virtual reality provides in its simplest implementation. Augmented reality, in turn, is a very revealing method of "expanding" this environment.

The use of gamification elements, interactive features, competition and self-control over the passage of the course are powerful tools for involvement in the learning process. E-education allows the student to gain the skills of teamwork, evaluation of other people's educational work, self-control. The e-learning system allows to automate many routine processes and get even more information about academic performance than with traditional education. The presence of an electronic version of the course is not only convenient for the student and teacher, but also shows the interest of teachers and students in global trends in education [3].

After our analysis of the currently available opportunities and the market for ready-made IT solutions, it was found that in most existing Learning Management Systems (LMS), such as Moodle, it is possible to create courses that require only text, graphic, as well as multimedia (audio and video fragments) information.

To analyze current IT solutions, some of the tools often mentioned in scientific and practical publications and currently used in practice for creating virtual and augmented reality, such as: Google Tilt Brush, Gravity Sketch, Oculus Medium, Varwin and the A-Frame framework (Table 1).

Table 1. Comparison of IT platforms for the implementation of VR/AR

| Software | VR support | AR support | Moodle integration | Visual programming support |
|---|---|---|---|---|
| Google Tilt Brush | + | + | - | - |
| Gravity Sketch | + | + | - | - |
| Oculus Medium | + | + | - | - |
| A-Frame | + | + | - | - |
| Varwin | + | + | - | + |

The main objective of our work is to create an easy-to-use constructor, which will be further integrated into the Moodle LMS and, as a result, will have all the necessary functions and a sufficient number of templates for the implementation of online training courses unprepared for the intricacies of developing VR/AR-content users, mainly teachers and course developers.

The key function of the tool we are developing is integration into the Moodle distance learning system, which is absent in all analogues presented for comparison. The second important feature - the possibility of visual programming - is present only in the Varwin platform, but this platform does not have the ability to integrate with the Moodle system. All of the above platforms have another important drawback - the complexity of use. So, for example, using A-Frame requires programming skills, and Varwin has a complex interface for an unprepared user, even its visual programming system requires certain skills.

Despite the fact that there are ready-made tools in the information products market for creating VR simulations and interactive educational applications, there is no universal approach for building final solutions or ready-made educational products. In this regard, it is not yet possible to develop a common conceptual methodology for creating VR-learning materials. Also, there are no flexible 2D/3D VR/AR construction sets on the market for different target groups of teachers or teachers (middle/high school, full or partial interactive, humanities or "techies", etc.). In addition, at the moment, all such software products are aimed at a specific user segment (in order to create social applications or computer games), and therefore there is no direct promotion in the field of educational services, and this marketing niche is also not busy. Another no less important reason for the development of a constructor module is the lack of Russian developments in this area.

In the process of studying an electronic educational course, there is no possibility of practical mastering of the material, students have to understand the topic of the lecture only on the basis of the lecture material. For teachers, the disadvantage is that in many cases it is impossible to explain in words a particular practical process, or the explanation process is very complex and the information is too difficult to understand, especially in technical specialties, although humanities specialties also often need a good example or immersion in a real work environment.

Based on the conducted research, we have identified the following critical requirements for the developed software:
- The system should inform the user about incorrect input of information;
- All messages must meet the requirements of clarity and logic;

- The choice of available 3D models and scenes should be implemented visually, each available model or scene should be represented as a miniature;
- Available 3D models and scenes should be grouped (e.g., by subject area);
- All available work scenarios must contain a detailed description;
- The script editing function must be available;
- The program must have a "friendly" and understandable interface for an unprepared user.

The software product is built using standardized and efficiently maintained solutions and is implemented as an open system, and allows for increasing functionality [4].

Table 2 lists the interoperability mechanisms implemented and the corresponding hardware support.

Table 2. Mechanisms of interaction

| Interaction mechanism | How does it increase engagement? | Hardware and support |
|---|---|---|
| Eye tracking | Gaze control (most users will need this) | VR headset like Cardboard |
| Traditional controllers | Works with widely supported Gamepad APIs | Can be used wherever the API is supported |
| Rotation controllers | Laser pointer effect | Can be used in Daydream and Gear VR devices |
| Position and rotation controllers | Fully trackable hands | Most VR devices like HTC Vive and Oculus Touch |
| Gesture tracking | Allows users to interact with the scene using gestures | e.g., Hololens |

In most cases, courses will be viewable on smartphones and Google Cardboard-like devices, so eye tracking will be the primary control.

Figure 1 shows a scheme that describes the functional cycle of a VR application. The headset provides position and rotation data; the developer uses this data to render the scene from the user's perspective, and then sends the rendered data to the headset, where it is distorted appropriately and displayed to the user.

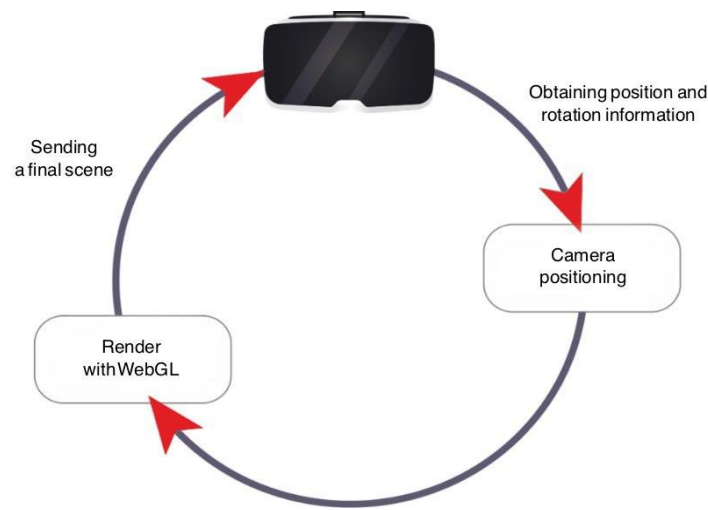

Figure 1. Scheme of the VR application work cycle

The developed module must be compatible with the Moodle e-learning system.

In order to develop a Moodle-compatible VR/AR constructor module, it may be need to define it as an Activity plugin. Plugins of this type, when integrated into Moodle, are simply placed in the appropriate folder and unpacked by the administrator.

The e-course organization chart in terms of development and functional content is as follows (Figure 2):

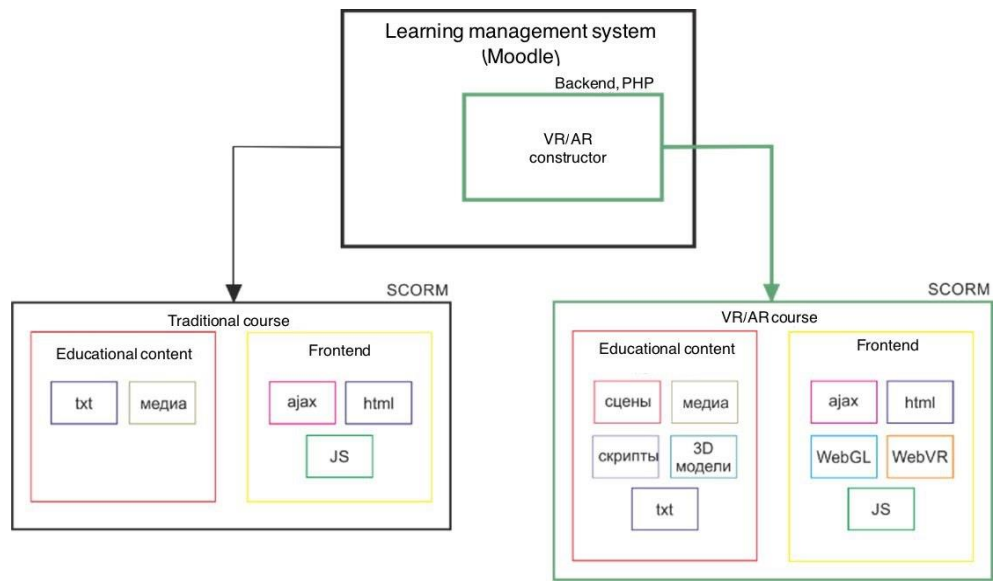

Figure 2. E-course organization chart

Creating VR/AR content is a very resource-intensive process. Combining and synchronizing the real world and user movements with the digital world requires a large number of graphic rendering processes. Because graphics are render intensive, on-device processes are complemented by workload sharing between the AR/VR device and the cloud (or server). Visualization of graphics in the cloud increases the sensitivity to delays when tracking the controller. This concept is called split rendering. But when rendering is done in the cloud rather than on a mobile device or desktop PC, a fast and reliable connection is essential to provide a seamless user experience. Based on this, the VR/AR-module designer must support work with 5G networks with further development.

Thus, in the course of this work, we have formulated the necessary functional and technological requirements for the development of a VR/AR-module-constructor of electronic educational courses. In addition, requirements were formulated for the interface, the appearance of the module-designer being developed in terms of its use by unprepared users. During the first stage of the software development project, the system architecture, a prototype of the graphical interface, template behavioral scripts and the object diagram of the repository of design blanks and 3D models were designed, and the software basis for the Moodle LMS Activity plugin was created.